\def\tr{{\text{tr}}\,}

\def\Im{{\text{Im}}\,}

\def\kF{k_{\text{F}}}
\def\vF{v_{\text{F}}}
\def\NF{N_{\text{F}}}

\def\sgn{{\text{sgn\,}}}
\def\qslash{{q\hskip -5pt /}}
\def\be{\begin{equation}}
\def\ee{\end{equation}}
\def\bea{\begin{eqnarray}}
\def\eea{\end{eqnarray}}
\def\bse{\begin{subequations}}
\def\ese{\end{subequations}}

\def\qslash{q\!\!\!/}
\documentclass[prb,twocolumn,showpacs,amsmath,amssymb,eqsecnum,preprintnumbers]{revtex4}
\usepackage{graphicx} 
\usepackage{dcolumn} 
\usepackage{bm} 
\usepackage{amsfonts}

\draft
\begin{document}
\title{Universal low-temperature tricritical point in metallic ferromagnets and ferrimagnets}
\author{T. R. Kirkpatrick$^1$, D. Belitz$^{2,3}$}
\affiliation{$^{1}$ Institute for Physical Science and Technology,and Department of 
                             Physics, University of Maryland, College Park, MD 20742, USA\\
$^{2}$ Department of Physics and Institute of Theoretical Science,
           University of Oregon, Eugene, OR 97403, USA\\
$^{3}$ Materials Science Institute, University of Oregon, Eugene, OR 97403, USA\\
}
\date{\today}

\begin{abstract}
An earlier theory of the quantum phase transition in metallic ferromagnets is
revisited and generalized in three ways. It is shown that the mechanism that 
leads to a fluctuation-induced first-order transition in metallic ferromagnets 
with a low Curie temperature
is valid, (1) irrespective of whether the magnetic moments are supplied by the
conduction electrons or by electrons in another band, (2) for ferromagnets in 
the XY and Ising universality classes as well as for Heisenberg ferromagnets,
and (3) for ferrimagnets as well as for ferromagnets. This vastly expands the
class of materials for which a first-order transition at low temperatures is
expected, and it explains why strongly anisotropic ferromagnets, such as
UGe$_2$, display a first-order transition as well as Heisenberg magnets.
\end{abstract}
\maketitle

\section{Introduction, and Results}
\label{sec:I}
Quantum phase transitions are a subject of great interest.\cite{Hertz_1976, Sachdev_1999}
In contrast to classical or thermal phase transitions, which occur
at a nonzero temperature $T_{\text c}>0$ and are driven by thermal fluctuations, quantum
phase transitions occur at zero temperature, $T=0$, as a function of some
non-thermal control parameter and are driven by quantum fluctuations. 
In this paper we will focus on quantum phase transitions
in metallic systems. For reasons discussed below, these transitions
are especially interesting.

A prototypical quantum phase transition is the one from a paramagnetic
metal to a ferromagnetic metal. Indeed, the earliest theory of a quantum
phase transition was the Stoner theory of ferromagnetism.\cite{Stoner_1938}
Stoner assumed that the conduction electrons are responsible for the
ferromagnetism, and developed a mean-field theory that describes both
the classical and the quantum ferromagnetic transition. 
In an important paper, Hertz later derived a Landau-Ginzburg-Wilson
(LGW) functional for this transition by considering a simple model
of itinerant electrons that interact only via a contact potential in the particle-hole
spin-triplet channel.\cite{Hertz_1976} Hertz analyzed this (dynamical) 
LGW functional by means of renormalization-group (RG) methods. He 
concluded that the critical behavior in the physical dimensions $d=2$ 
and $d=3$ is mean-field-like. That is, as far as the static critical exponents 
of the transition at $T=0$ are concerned, he concluded that Stoner theory is 
exact in $d=2$ and $d=3$.

In the mid 1990s it was realized that the above conclusion is not correct.
The problem is that in metals at $T=0$ there are gapless particle-hole
excitations that couple to the magnetic order-parameter fluctuations
and influence the quantum critical behavior for all dimensions $d \leq 3$.
In Hertz's theory this coupling is taken into account only in an approximation
that does not suffice for yielding the leading critical behavior. Technically,
Hertz theory treats the fermionic soft modes in a tree approximation,
whereas describing their influence on the critical behavior requires taking
into account fermionic loops. Physically, a correct description of any
phase transition must treat the order parameter fluctuations and all soft 
modes that couple to them on equal footing.

\begin{figure}[t]
\vskip -0mm
\includegraphics[width=8.0cm]{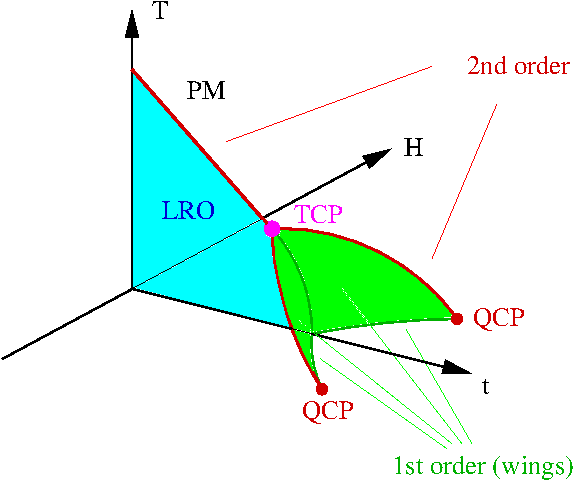}
\caption{Generic phase diagram of a metallic magnet in the space spanned by temperature ($T$),
               magnetic field ($H$), and the control parameter ($t$). Shown are the
               long-range ordered magnetic (LRO) and paramagnetic (PM) phases, 
               lines of second-order transitions, surfaces of first-order transitions 
               (``tricritical wings''), the tricritical point (TCP), and the two quantum 
               critical points (QCP). The long-range order can be of ferromagnetic or
               ferrimagnetic type, and the electrons causing the long-range order
               can be in the same band as the conduction electrons, or in a different
               band. See the text for further explanation.}
\label{fig:1}
\end{figure}

A theory that takes into account these effects was developed by the present
authors and T. Vojta. In Ref.\ \onlinecite{Belitz_Kirkpatrick_Vojta_1999} it
was shown that the quantum phase transition from a metallic paramagnet
to an itinerant ferromagnet in the absence of quenched disorder
in $d=2$ and $d=3$ is generically discontinuous, 
or of first order, in contrast to the second-order transition with mean-field 
critical behavior predicted by Hertz theory.\cite{dimensionality_footnote}  
The mechanism behind this phenomenon is analogous to what is known
as a fluctuation-induced first-order transition in superconductors and
liquid crystals.\cite{Halperin_Lubensky_Ma_1974} There, soft fluctuations
of the electromagnetic vector potential (in superconductors) or the nematic
order parameter (in liquid crystals) couple to the order parameter and effectively change
the sign of the cubic term in the equation of state, leading to a first-order
transition. In the quantum magnetic case, the role of the additional soft
modes is played by the fermionic particle-hole excitations mentioned
above that are massless at $T=0$. Since these modes acquire a mass at 
$T>0$, the tendency towards a first-order transition diminishes with 
increasing temperature. This leads to a tricritical point at a temperature
$T_{\text{tc}}>0$ that separates a line of continuous transitions at
$T > T_{\text{tc}}$ from a line of first-order transitions at $T < T_{\text{tc}}$.
In a later paper with Rollb{\"u}hler, the effects of a magnetic field $H$ were
investigated.\cite{Belitz_Kirkpatrick_Rollbuehler_2005} It was found that
in the space spanned by $T$, $H$, and the control parameter, tricritical
wings, or surfaces of first-order transitions, emanate from the tricritical
point and terminate in a pair of quantum critical points in the $T=0$
plane. The wing boundaries at $T>0$ are given by lines of critical points 
that are reminiscent of a conventional liquid-gas critical point and connect
the tricritical point with the quantum critical points at $T=0$. The resulting
generic phase diagram is shown in Fig.\ \ref{fig:1}. 
This general picture is in good agreement with experimental results for
low-Curie-temperature metallic ferromagnets, including 
ZrZn$_2$,\cite{Uhlarz_Pfleiderer_Hayden_2004} UGe$_2$,\cite{Taufour_et_al_2010}
URhGe,\cite{Yelland_et_al_2011}
and MnSi.\cite{Pfleiderer_Julian_Lonzarich_2001, MnSi_footnote}

In this paper we generalize our previous theory in three important
ways. First, we show that our previous results, which had been derived under
the same assumption made by Stoner and by Hertz, namely, 
that the magnetism is caused only by itinerant electrons, remain valid
in metallic systems where the magnetism is caused by electrons in a
different band than the conduction electrons. 

Second, we show that the results are {\em not} restricted to Heisenberg
ferromagnets, contrary to what was implied in Refs.\ \onlinecite{Belitz_Kirkpatrick_Vojta_1999}
and \onlinecite{Kirkpatrick_Belitz_2003b}. Rather, they apply equally well to
metallic XY or Ising magnets, since the magnetic moments couple to
conduction electrons whose spins have three degrees of freedom. This is
an important point, since some of the relevant materials are strongly
anisotropic magnets, including UGe$_2$ (easy axis) and URhGe (easy plane).

Third, we show that the phase diagram shown in Fig.\ \ref{fig:1} also applies
to generic metallic ferrimagnets. Ferrimagnets are materials that
spontaneously develop both a homogeneous and a staggered magnetization
at the same critical value of either the temperature (for a classical transition)
or a non-thermal control parameter (for a quantum transition).
Physically, this can happen when magnetic moments of unequal magnitude 
on a bipartite lattice align in opposite directions.\cite{Kittel_1996}

The unifying principle behind these generalizations is the realization that 
coupling a homogeneous magnetization to conduction electrons will  
produce the same results irrespective of the microscopic origin of the
magetization.\cite{MFT_footnote} As a result, the phase diagram depicted
schematically in Fig.\ \ref{fig:1} is valid for generic metallic 
ferromagnets in addition to itinerant ones, for ferromagnets of XY or Ising type
in addition to Heisenberg magnets, and for ferrimagnets as well as for
ferromagnets. In all cases we also consider the effects of nonmagnetic quenched
disorder. In Ref.\ \onlinecite{Belitz_Kirkpatrick_Vojta_1999} it was shown 
that this type of disorder leads to an interesting phase diagram with a
number of multi-critical points, and that sufficiently strong quenched disorder
causes the first-order paramagnetic-to-ferromagnetic transition in metals to become 
second order. We will see that the same result holds for metallic ferrimagnets.
Experimentally, the effects of disorder on either one of these transitions have 
not yet been studied systematically.

\section{Theory}
\label{sec:II}

We now derive the results listed in Sec.\ \ref{sec:I}. To this end we are interested in a
theory that describes the magnetization or order-parameter (OP) field ${\bm M}$, the
fermionic degrees of freedom described by Grassmann-valued fields $\bar\psi$ and
$\psi$, and the coupling between them. Accordingly, the action will have three parts:
\bse
\label{eqs:2.1}
\be
{\cal A}[{\bm M};{\bar\psi},\psi] = {\cal A}_{\text{OP}}[{\bm M}] + {\tilde{\cal A}}_{\text{F}}[\bar\psi,\psi] 
   + {\tilde{\cal A}}_{\text{c}}[{\bm M};\bar\psi,\psi]\ ,
\label{eq:2.1a}
\ee
and the partition function is given by
\be
Z = \int D[{\bm M}]\,D[{\bar\psi},\psi]\, e^{-{\cal A}[{\bm M};{\bar\psi},\psi]}\ .
\label{eq:2.1b}
\ee
\ese
We are, however, not interested in a complete description of the fermionic degrees
of freedom; rather, we want to restrict ourselves to the fermionic soft modes and
integrate out the massive modes in the simplest approximation that respects the
symmetries of the problem to arrive at an effective Landau-Ginzburg-Wilson (LGW)
theory in terms of soft modes only. If we denote the soft fermionic degrees of 
freedom collectively by $q$, and the massive ones by $P$, we formally have
\bse
\label{eqs:2.2}
\be
Z = \int D[{\bm M},q]\ e^{-{\cal A}_{\text{LGW}}[{\bm M},q]}\ ,
\label{eq:2.2a}
\ee
where
\bea
{\cal A}_{\text{LGW}}[{\bm M},q] &=& {\cal A}_{\text{OP}}[{\bm M}] 
   - \ln \int D[P]\ e^{-{\tilde{\cal A}}_{\text{F}}[q,P]}
\nonumber\\
&&\hskip 90pt \times e^{ - {\tilde{\cal A}}_{\text{c}}[{\bm M};q,P]}
\nonumber\\
&\equiv& {\cal A}_{\text{OP}}[{\bm M}] + {\cal A}_{\text{F}}[q] +  {\cal A}_{\text{c}}[{\bm M},q]\ .
\label{eq:2.2b}
\eea
\ese
As we will see later, the $q$ are matrices formed by bilinear products of the fermion fields, 
$q_{nm}({\bm x},{\bm y}) = {\bar\psi}_n({\bm x})\,\psi_m({\bm y})$ with $(n+1/2)(m+1/2)<0$,
and the $P$ are given by the same products with $(n+1/2)(m+1/2)>0$. 
Here $\psi_n({\bm x}) \equiv \psi({\bm x},\omega_n)$ is the temporal Fourier transform of the
Grassmann field $\psi(x)$, where $x \equiv ({\bm x},\tau)$ comprises the real-space position
${\bm x}$ and the imaginary-time variable $\tau$ in a Matsubara formalism, and $\omega_n =
2\pi T(n+1/2)$ is a fermionic Matsubara frequency. ${\bar\psi}_n({\bm x})$ is defined analogously.

This  separation of soft and massive fermionic modes $q$ and $P$, respectively, integrating out 
$P$ in a suitable approximation, and determining the consequences of the coupling between
$q$ and ${\bm M}$, is the central objective of this paper. For the separation we will make use
of the general theory developed in Ref.\ \onlinecite{Belitz_Kirkpatrick_2012}.

\subsection{Order parameter, and coupling to fermions}
\label{subsec:II.A}

We are interested in magnetic order, and hence the appropriate order-parameter field
is the magnetization ${\bm M}(x)$.
We write the magnetization as a part ${\bm m}(x)$ whose average is the homogeneous
magnetization, and a part ${\bm n}(x)$ whose average is a staggered magnetization,
\be
{\bm M}(x) = {\bm m}(x) + {\bm n}(x)\,\sum_{j=1}^{N}\cos({\bm k}_j\cdot{\bm x})\ .
\label{eq:2.3}
\ee
Here the ${\bm k}_j$ are $N$ wave vectors that characterize the staggered magnetic
order, and both ${\bm m}(x)$ and ${\bm n}(x)$ are slowly varying in space and time.
In particular, their Fourier expansions contain only wave numbers that are small compared
to the norms of the ${\bm k}_j$. 

In a paramagnetic state the expectation values of ${\bm m}$ and ${\bm n}$ are both zero.
At a transition to a ferromagnetic state the expectation value of ${\bm m}$ becomes
nonzero while that of ${\bm n}$ remains zero; at a transition to an antiferromagnetic state
the converse is true. A ferrimagnetic transition is characterized by both ${\bm m}$ and
${\bm n}$ acquiring a nonzero expectation value at the same point in parameter space.
In this sense there is only one order parameter field for a ferrimagnetic transition; this fact
will be importan later. For the purposes of the present paper, a crucial question is the 
coupling of the order-parameter fluctuations to the soft fermionic degrees of freedom. 
Since the soft parts of the latter are soft at zero wave number,
the leading coupling is to ${\bm m}$. The fermions also couple to ${\bm n}$, but this
leads to subleading effects since the staggered magnetization is soft at a nonzero
wave number. We will neglect this coupling in what follows. 
Physically, the near-homogeneous magnetization flutuations act as a magnetic field
proportional to ${\bm m}$ that couples to the electronic spin density 
\bse
\label{eqs:2.4}
\be
{\bm n}_{\text{s}}(x) = \sum_{a,b}{\bar\psi}_a(x)\,{\bm\sigma}_{ab}\,\psi_b(x)\ .
\label{eq:2.4a}
\ee
Here ${\bm\sigma} = (\sigma^x,\sigma^y,\sigma^z) \equiv (\sigma^1,\sigma^2,\sigma^3)$ 
denotes the Pauli matrices,
and $a,b = (\uparrow,\downarrow) \equiv (+1,-1)$ are spin indices. The coupling
takes the form of a Zeeman term
\be
\tilde{\cal A}_{\text{c}}[M;\bar\psi,\psi] = c \int dx\ {\bm m}(x)\cdot{\bm n}_{\text{s}}(x)\ ,
\label{eq:2.4b}
\ee
\ese
with $c$ a coupling constant. As we will see, the spin density contains both massive and
massless modes, so only part of Eq.\ (\ref{eq:2.4b}) contributes to ${\cal A}_{\text{c}}[{\bm M},q]$
in Eq.\ (\ref{eq:2.2b}). We will discuss this separation next.

\subsection{Fermionic soft modes}
\label{subsec:II.B}

In this subsection we separate the massless fermionic modes from the massive ones by
means of the technical apparatus developed in Ref.\ \onlinecite{Belitz_Kirkpatrick_2012}.
Here we will quote only as much of this formalism as is necessary for the further
development, see Ref.\ \onlinecite{Belitz_Kirkpatrick_2012} for additional details.

The soft fermion excitations are all two-particle excitations; the related
correlation functions are those of bilinear products of fermion fields. The latter commute with
each other, and with individual fermion fields, and hence are isomorphic to classical
fields. Denoting these classical fields by $Q$, we define a classical matrix field
\begin{widetext}
\be
Q_{nm}({\bm x},{\bm y}) \cong \frac{i}{2}\left(\begin{array}{cccc}
                -\psi_{n\uparrow}({\bm x}){\bar\psi}_{m\uparrow}({\bm y}) & -\psi_{n\uparrow}({\bm x}){\bar\psi}_{m\downarrow}({\bm y})
                       & -\psi_{n\uparrow}({\bm x}) \psi_{m\downarrow}({\bm y}) & \ \ \psi_{n\uparrow}({\bm x}) \psi_{m\uparrow}({\bm y}) \\
               -\psi_{n\downarrow}({\bm x}){\bar\psi}_{m\uparrow}({\bm y}) & -\psi_{n\downarrow}({\bm x}){\bar\psi}_{m\downarrow}({\bm y})
                       & -\psi_{n\downarrow}({\bm x})\psi_{m\downarrow}({\bm y}) & \ \ \psi_{n\downarrow}({\bm x})\psi_{m\uparrow}({\bm y}) \\
                 \ \ {\bar\psi}_{n\downarrow}({\bm x}){\bar\psi}_{m\uparrow}({\bm y}) & \ \ {\bar\psi}_{n\downarrow}({\bm x}){\bar\psi}_{m\downarrow}({\bm y})
                      & \ \ {\bar\psi}_{n\downarrow}({\bm x})\psi_{m\downarrow}({\bm y}) & - {\bar\psi}_{n\downarrow}({\bm x})\psi_{m\uparrow}({\bm y})\\
                - {\bar\psi}_{n\uparrow}({\bm x}){\bar\psi}_{m\uparrow}({\bm y}) & -{\bar\psi}_{n\uparrow}({\bm x}){\bar\psi}_{m\downarrow}({\bm y})
                      & -{\bar\psi}_{n\uparrow}({\bm x})\psi_{m\downarrow}({\bm y}) & \ \ {\bar\psi}_{n\uparrow}({\bm x})\psi_{m\uparrow}({\bm y})
                  \end{array}\right)\ .
\label{eq:2.5}
\ee
\end{widetext}
Here ``$\cong$'' means ``isomorphic to''; technically, the isomorphism is implemented by
means of a Lagrange multiplier field, see below. We also define
the Fourier transform of $Q$,
\bse
\label{eqs:2.6}
\be
Q_{nm}({\bm k},{\bm p})=\frac{1}{V}\int d{\bm x}\,d{\bm y}\ e^{-i{\bm k}\cdot{\bm x} 
   +i{\bm p}\cdot{\bm y}}\,Q_{nm}({\bm x},{\bm y})\ .
\label{eq:2.6a}
\ee
It is further useful to define
\be
Q_{nm}({\bm k};{\bm q}) = Q_{nm}({\bm k}+{\bm q}/2,{\bm k}-{\bm q}/2)
\label{eq:2.6b}
\ee
and
\be
Q_{nm}({\bm x}) = Q_{nm}({\bm x},{\bm x}) = \frac{1}{V}\sum_{\bm q} e^{i{\bm q}\cdot{\bm x}}
   \sum_{\bm k} Q_{nm}({\bm k};{\bm q})\ .
\label{eq:2.6c}
\ee
\ese
The $4\times 4$ matrix $Q_{nm}$ can be expanded in a spin-quaternion basis
\be
Q_{nm}({\bm x},{\bm y}) = \sum_{r,i=0}^3 (\tau_r\otimes s_i)\,{^i_r Q}_{nm}({\bm x},{\bm y})\ ,
\label{eq:2.7}
\ee
where $\tau_0 = s_0 = \openone_2$ is the unit $2\times 2$ matrix, and $\tau_{1,2,3} = -s_{1,2,3}
= -i\sigma^{1,2,3}$. An explicit inspection of the 16 matrix elements shows that $r=0,3$
represents the particle-hole channel, i.e., products of the form ${\bar\psi}\psi$, whereas $r=1,2$
represents the particle-particle channel, i.e., products of the form ${\bar\psi}{\bar\psi}$ or
$\psi\psi$. For our purposes we will need only the particle-hole degrees of freedom.

It was shown in Ref.\ \onlinecite{Belitz_Kirkpatrick_2012} (see also 
Ref.\ \onlinecite{Belitz_Kirkpatrick_1997}) that a crucial criterion for separating the fermionic
degrees of freedom into soft and massive modes is given by the relative signs of the frequency arguments
of the matrix elements $Q_{nm}$. Accordingly, we write
\bea
{^i_r Q}_{nm}({\bm x}) &=& {^i_r q}_{nm}({\bm x})\,\Theta(-\omega_n\omega_m) +
     {^i_r P}_{nm}({\bm x})\,\Theta(\omega_n\omega_m)
\nonumber\\
&& \hskip 90pt (i=1,2,3)
\label{eq:2.8}
\eea
Here $\Theta$ is the step function, and we use the fact that in the spin-triplet channel 
($i=1,2,3$) the expectation value of the $Q$-matrix vanishes (this is since the fermionic
degrees of freedom described by $Q$ do not by themselves have long-ranged magnetic
order; see the discussion at the end of the current subsection), 
so that $q$ and $P$ represent fluctuations. In what follows we will absorb the step functions
into the matrix fields $q$ and $P$, i.e., writing $q_{nm}$ implies $n \geq 0$ and $m<0$ and
$P_{nm}$ implies either $n\geq 0$ and $m\geq 0$ or $n<0$ and $m<0$. The ${^i_r q}$ are
the spin-quaternion elements of a matrix
\bse
\label{eqs:2.9}
\be
q_{nm}({\bm x}) = \sum_{r,i} (\tau_r\otimes s_i)\,{^i_r q}_{nm}({\bm x})\ .
\label{eq:2.9a}
\ee
It is also useful to define an adjoint matrix
\be
q^+_{nm}({\bm x}) = \sum_{i,r} (\tau_r^+\otimes s_i^+)\,{^i_r q}_{mn}({\bm x})\ ,
\label{eq:2.9b}
\ee
\ese
where $\tau_r^+$ and $s_i^+$ are the hermitian conjugates of $\tau_r$ and $s_i$, respectively.
In addition, the theory contains a field $\qslash_{nm}({\bm x})$ that has the same properties as
$q_{nm}({\bm x})$ except for different propagators, see below. The origin of $\qslash$ is the
Lagrange multiplier field $\lambda$ that constrains the bilinear products of fermion fields to the $q$.
In various places in the theory $q - \lambda \equiv \qslash$ appears, and the $\lambda$-propagator
equals minus the $q$-propagator for noninteracting electrons, whereas cross-correlations between
$q$ and $\lambda$ vanish. The net effect of $\lambda$ is therefore to subtract the noninteracting 
part of the $q$-propagator wherever the combination $q-\lambda$ occurs.

The $q$ correlation functions are the basic soft modes in the theory, see below. However,
due to nonlinear couplings the $P$ couple to the $q$ and thus have a soft component. This
effect can be expressed by expanding $P$ in a power series in $q$. To quadratic order in $q$
and to lowest order in the fermion interaction one finds
\bea
P_{12}({\bm k}) &\approx& -2i\sum_3\sum_{\bm p}\varphi^{(3)}_{132}({\bm p},{\bm k}-{\bm p})\,\varphi^{-1}_{13}({\bm p})\,
     \varphi^{-1}_{32}({\bm k}-{\bm p})
\nonumber\\
&& \hskip -20pt \times \left[\qslash_{13}({\bm p})\,\qslash^+_{32}({\bm k}-{\bm p}) 
     + \qslash^+_{13}({\bm p})\,\qslash_{32}({\bm k}-{\bm p})\right]\ .
\label{eq:2.10}
\eea
Here and it what follows we use a simplified notation for frequency indices, $1\equiv n_1$, etc.
We have dropped contributions to $P$ of higher order in $q$, and a contribution that is linear in the interaction
and linear in $q$, see Ref.\ \onlinecite{Belitz_Kirkpatrick_2012}; neither will be needed for our purposes.
We also have omitted a term quadratic in $q$ and quadratic in the interaction, which leads to less singular
contributions to the free energy than the one we keep.
Note the frequency restrictions inherent in Eq.\ (\ref{eq:2.10}): $\sgn(\omega_{n_1}) = \sgn(\omega_{n_2}) = -\sgn(\omega_{n_3})$.
Here
\begin{equation}
\varphi_{12}({\bm k}) = \frac{1}{V}\sum_{\bm p} G_1({\bm p})\,G_2({\bm p}-{\bm k})
\label{eq:2.11}
\end{equation}
with $\omega_{n_1}\,\omega_{n_2} < 0$ implied, and 
\be
\varphi_{132}^{(3)}({\bm k}_1,{\bm k}_2)  = \frac{1}{V}\sum_{\bm p} G_1({\bm p})\,G_3({\bm p}-{\bm k}_1)\,
      G_2({\bm p}-{\bm k}_1-{\bm k}_2)
\label{eq:2.12}
\ee
where $G_1({\bm p}) \equiv G({\bm p},i\omega_{n_1})$ is the single-particle Green function. 
$\varphi_{12}$ has a scaling form
\bea
\varphi_{12}({\bm k}) &=& \NF\,\frac{2\pi G}{k}\,\varphi_d(Gi\Omega_{1-2}/k)
\nonumber\\
&\equiv& \varphi({\bm k},\Omega_{1-2})\ .
\label{eq:2.13}
\eea
where $G$ is a coupling constant whose bare value is the inverse Fermi velocity, $G = 1/\vF$, $\NF$ is the
density of states per spin at the Fermi level, and $\Omega_{1-2} = \omega_{n_1} - \omega_{n_2}$. In $d=2,3$,
and for free electrons, we find explicitly
\bse
\label{eqs:2.14}
\bea
\varphi_{d=2}(z) &=& \sgn(\Im z)/\sqrt{1-z^2}\ ,
\label{eq:2.14a}\\
\varphi_{d=3}(z) &=& \frac{-i}{2}\,\ln\left(\frac{1-z}{-1-z}\right)\ ,
\label{eq:2.14b}
\eea
\ese
which we recognize as the hydrodynamic part of the Lindhard function. Equations (\ref{eq:2.13})
and (\ref{eqs:2.14}) reflect the soft particle-hole excitations with a linear momentum-frequency
relation in a metallic electron system. In particular, $\varphi({\bm k},\Omega_n=0) \propto 1/\vert{\bm k}\vert$,
and $\varphi({\bm k}=0,\Omega_n) \propto 1/\Omega_n$.\cite{propagator_footnote} 
For later reference we also note
the following identities that hold for a special form of $\varphi^{(3)}$:
\bea
\varphi^{(3)}_{121}({\bm k},-{\bm k}) &=&  -\varphi^{(3)}_{212}({\bm k},-{\bm k}) 
   = -\,\frac{\partial}{\partial i\omega_{n_1}}\,\varphi_{12}({\bm k})
\nonumber\\
&\equiv& \varphi^{(3)}({\bm k},\Omega_{1-2})\ .
\label{eq:2.15}
\eea

The fermionic action can be expressed in terms of $q$ and $P$, and by using Eq.\ (\ref{eq:2.10}) and
its generalizations to higher order one obtains a fermionic soft-mode action entire in terms of $q$.
For our purposes we need only the Gaussian part of this action, which reads
\bse
\label{eqs:2.16}
\be
{\cal A}_{\text F}[q] = -8 \sum_{\bm k}\sum_{1,2\atop 3,4}\sum_{r=0,3} \sum_{i=0}^3 
   {^i_r q}_{12}({\bm k})\ \Gamma^i_{12,34}({\bm k})\, {^i_r q}_{34}(-{\bm k})\ .
\label{eq:2.16a}
\ee
Here $1\equiv n_1$ etc., and the Gaussian vertex is given by
\be
\Gamma^i_{12,34}({\bm k}) = \varphi_{12}^{-1}({\bm k}) + \delta_{1-2,3-4}\,2T\gamma^i
\label{eq:2.16b}
\ee
\ese
with $\gamma^{i=0} = -\gamma_{\text{s}}$ and $\gamma^{i=1,2,3} = \gamma_{\text{t},i}$,
where $\gamma_{\text{s}}>0$ and $\gamma_{\text{t},i} > 0$ are the spin-singlet and 
spin-triplet interaction amplitudes. 
The fermionic Gaussian propagator is given by the inverse of the vertex. One finds
\begin{widetext}
\bse
\label{eqs:2.17}
\be
\langle{^i_r q}_{12}({\bm k})\,{^j_s q}_{34}(-{\bm k})\rangle = \frac{1}{16}\,\delta_{rs}\,\delta_{ij}
   \left[\delta_{13}\,\delta_{24}\,\varphi_{12}({\bm k})
     - 2\gamma^i T\,\delta_{1-2,3-4}\,\frac{\varphi_{12}({\bm k})\,\varphi_{34}({\bm k})}
           {1 - 2\gamma^i\chi^{(0)}_{1-2}({\bm k})}\right]\ ,
\label{eq:2.17a}
\ee
where
\be
\chi^{(0)}_{1-2}({\bm k}) \equiv \chi^{(0)}({\bm k},\Omega_{1-2}) = -T\sum_{34}\delta_{1-2,3-4}\,\varphi_{34}({\bm k})\ .
\label{eq:2.17b}
\ee
We see that the $q$-propagator is given in terms of $\varphi$, and hence is soft.
The fields $\qslash$ that enter $P$, Eq.\ (\ref{eq:2.10}), are characterized by Gaussian
propagators
\be
\langle{^i_r \qslash}_{12}({\bm k})\,{^j_s q}_{34}(-{\bm k})\rangle = \langle{^i_r q}_{12}({\bm k})\,{^j_s \qslash}_{34}(-{\bm k})\rangle 
   = \langle{^i_r q}_{12}({\bm k})\,{^j_s q}_{34}(-{\bm k})\rangle 
\label{eq:2.17c}
\ee
and
\be
\langle{^i_r \qslash}_{12}({\bm k})\,{^j_s \qslash}_{34}(-{\bm k})\rangle 
        =  \frac{-1}{8}\,\gamma^i T\,\delta_{1-2,3-4}\,\frac{\varphi_{12}({\bm k})\,\varphi_{34}({\bm k})}
           {1 - 2\gamma^i\chi^{(0)}_{1-2}({\bm k})}\ .
\label{eq:2.17d}
\ee
\ese
\end{widetext}
The last expression is just the interacting part of the $q$-propagator, Eq.\ (\ref{eq:2.17a}), as was
mentioned after Eq.\ (\ref{eq:2.9b}).

The interaction amplitudes in the Gaussian fermionic vertex, Eq.\ (\ref{eq:2.16b}), warrant some
comments. First, we note that the three spin-triplet amplitudes $\gamma_{\text{t}}^{1,2,3}$ 
are in general not identical in a cyrstalline solid, and they do not need to be for what follows. 
Second, we comment on the two cases that result from the magnetism being caused by the
conduction electrons, or by electrons in a band different from the conduction band, respectively.
Let us first assume the latter case, which is the conceptually more straightforward one. 
Then ${\cal A}_{\text{F}}[q]$, which
describes the conduction electrons, is independent of the magnetism and contains interactions in
both the spin-singlet and spin-triplet channels. The only restriction is that the latter are weak enough
to not lead to magnetism by themselves. The conduction electrons are affected by the magnetization,
which acts as an effective magnetic field, and this is described by the Zeeman coupling term, 
Eq.\ (\ref{eq:2.4b}). The other possibility, which is conceptually more complex, 
is that the magnetism is caused by the
conduction electrons themselves. In this case the magnetic order parameter and the soft modes $q$
describe degrees of freedom for electrons in the same band. The magnetic order parameter then should
be thought of as deriving from the spin-triplet interaction between the conduction electrons, e.g., via
a Hubbard-Stratonovich decoupling of the latter. This leaves the bare action ${\cal A}_{\text{F}}$ with a
spin-singlet interaction only. However, as long as the latter is present, a spin-triplet interaction will
always be generated under renormalization. The action ${\cal A}_{\text{F}}$ will therefore again contain
a spin-triplet interaction amplitude, albeit one that is much weaker than the one in the underlying
action that describes the system before the separation of magnetic and fermionic degrees of freedom.
This is the case that was discussed, for ferromagnetism, in Ref.\ \onlinecite{Kirkpatrick_Belitz_2003b},
which used phenomenological and symmetry arguments to construct the fermionic part of the action.
Finally, we mention that we assume the conduction electrons, in the absence of a nonzero
magnetization (i.e., with the coupling constant $c$ in Eq.\ (\ref{eq:2.4b}) put equal to zero), to indeed 
have three soft spin-triplet
excitations at $T=0$, which are given by Eqs.\ (\ref{eqs:2.17}) with $i=1,2,3$. This is not
necessarily the case. For instance, an external magnetic field gives two of these three channels
(the ones transverse to the field) a mass, and a small concentration of magnetic impurities will
make all three channels massive without having significant other effects.
However, in general the energy
scales associated with these effects will be small, and they will lead to a small reduction, but not
a complete suppression, of the tricritical temperature in Fig.\ \ref{fig:1}. We will discuss this point
in more detail in Sec.\ \ref{sec:III}.

\subsection{Coupling between the order parameter and the fermionic soft modes}
\label{subsec:II.C}

We are now in a position to separate the Zeeman term, Eq.\ (\ref{eq:2.4b}),
into parts where the order parameter couples to soft and massive fermionic modes, respectively.
If we define a temporal Fourier transform of the magnetization field $m$ by
\be
{\bm m}_n({\bm x}) = \sqrt{T} \int_0^{1/T} d\tau\ e^{i\Omega_n\tau}\,{\bm m}({\bm x},\tau)\ ,
\label{eq:2.18}
\ee
with $\Omega_n = 2\pi Tn$ a bosonic Matsubara frequency, then we can write Eq.\ (\ref{eq:2.4b})
in the form
\bea
{\tilde{\cal A}}_{\text{c}}[{\bm M};Q] &=& 2c\sqrt{T} \int d{\bm x} \sum_n \sum_{i=1}^3 m_n^i({\bm x})
\nonumber\\
&& \hskip -60pt \times \sum_{r=0,3} (-1)^{r/2} \sum_m \tr [(\tau_r \otimes s_i)\,Q_{m,m+n}({\bm x})]\ .
\label{eq:2.19}
\eea
By expressing $Q$ in terms of $q$ and $P$ by means of Eq.\ (\ref{eq:2.8}), and $P$ in terms of $\qslash$
by means of Eq.\ (\ref{eq:2.10}), we obtain the desired coupling ${\cal A}_{\text{c}}[{\bm M},q]$ between 
the order-parameter fluctuations and the fermionic soft modes $q$.

\subsection{Generalized Mean-Field Theory}
\label{subsec:II.D}

An effective action, ${\cal A}_{\text{eff}}[{\bm M}]$ in terms of the order parameter alone can be obtained
by integrating out the fields $q$,
\be
{\cal A}_{\text{eff}}[{\bm M}] = \ln \int D[q]\ e^{{\cal A}_{\text{LGW}}[{\bm M},q]}\ .
\label{eq:2.20}
\ee

In general the evaluation of this expression is very difficult. However,
it can be evaluated exactly within a generalized mean-field approximation that was first employed
in the context of liquid crystals and superconductors\cite{Halperin_Lubensky_Ma_1974} and is
defined as follows. First, we ignore temporal and spatial variations of the order parameter, i.e. we
treat the fields ${\bm m}(x)$ and ${\bm n}(x)$ in Eq.\ (\ref{eq:2.3}) as numbers. If we assume
ordering in the 3-direction, we have
\bse
\label{eqs:2.21}
\be
M^i(x) \approx \delta_{i3}\,\left[m + n\sum_{j=1}^N \cos({\bm k}_j\cdot{\bm x})\right]\ ,
\label{eq:2.21a}
\ee
which implies
\be
m_n^i(x) \approx \delta_{i3}\,\delta_{n0}\,m/\sqrt{T}\ .
\label{eq:2.21b}
\ee
\ese
This mean-field approximation for the order parameter means that only the part of $Q$
that is diagonal in frequency space, i.e., $P_{mm}$, contributes to Eq.\ (\ref{eq:2.19}). This 
in turn means
that the contribution to $P$ that is linear in $q$, which we had dropped from Eq.\ (\ref{eq:2.10}),
does not contribute. Second, we restrict ourselves to quadratic order in $q$. That is, we treat
the fermionic soft modes in a Gaussian approximation with a fixed magnetic order parameter.
The validity of these approximations will be discussed in Sec.\ \ref{subsec:III.B}.

With these approximations the action ${\cal A}_{\text{c}}$ that couples $q$ and the order
parameter is quadratic in $q$ and can be written
\begin{widetext}
\bse
\label{eqs:2.22}
\be
{\cal A}_{\text{c}}[m,q] = 8\sum_{r,s=0,3}\sum_{i,j} {^i_r q}_{12}({\bm k})\,
   {^{ij}_{rs}\Gamma}^{c}_{12,34}({\bm k})\,{^j_s q}_{34}(-{\bm k})\ .
\label{eq:2.22a}
\ee
Here
\be
 {^{ij}_{rs}\Gamma}^{c}_{12,34}({\bm k}) = \delta_{13}\,\delta_{24}\,4\,c\,m\left(\begin{array}{cc} 0 & 1 \\
                                                                                                          -1 & 0 \end{array}\right)_{rs}
    \left(\begin{array}{cccc} 0 & 0 & 0 & 0 \\
                                          0 & 0 & 1 & 0 \\
                                          0 &-1& 0 & 0 \\
                                          0 & 0 & 0 & 0 \end{array}\right)_{ij} \varphi^{(3)}_{121}({\bm k},-{\bm k})\,\varphi^{-2}_{12}({\bm k})\ ,
\label{eq:2.22b}
\ee
\ese
and we have used Eq.\ (\ref{eq:2.15}). The matrices give the values of ${^{ij}_{rs}\Gamma}^c$
for the 4 possible values of $(r,s)$ and the 16 possible values of $(i,j)$.

The integral over $q$ in Eq.\ (\ref{eq:2.20}) can now easily be carried out. For the
free-energy density $f = -T{\cal A}_{\text{eff}}/V$ we obtain
\bse
\label{eqs:2.23}
\be
f = f_{0}(m,n) + \Delta f(m)\ .
\label{eq:2.23a}
\ee
Here $f_0 = -T{\cal A}_{\text{OP}}/V$ is the mean-field free energy in the absence of a 
coupling to the fermionic soft modes. For $\Delta f(m)$, which is the contribution to the
free energy due to this coupling, one finds
\be
\Delta f(m) = \frac{2}{V}{\sum_{\bm k}}^{\prime} T\sum_n \ln N({\bm k},\Omega_n;m)\ ,
\label{eq:2.23b}
\ee
where $\sum_{\bm k}^{\prime}$ denotes a wave vector sum such
that $\vert{\bm k}\vert < \Lambda$ with $\Lambda$ an ultraviolet cutoff, and
\be
N({\bm k},\Omega_n;m) = -16\,c^2\,\gamma_{\text{t},1}\gamma_{\text{t},2}\,m^2\,\Omega_n^2\,
   \left(\varphi^{(3)}({\bm k},\Omega_n)\right)^2\varphi^{-4}({\bm k},\Omega_n) 
+ \varphi^{-4}({\bm k},\Omega_n)\prod_{i=1,2}
     \left[1 - 2\gamma_{\text{t},i} \chi^{(0)}({\bm k},\Omega_n)\right]\ .
\label{eq:2.23c}
\ee
\ese
\end{widetext}

The equation of state is obtained by minimizing the free energy density. In the absence of a coupling
between the order parameter and the fermionic soft modes this amounts to minimizing $f_0$, which
yields the ordinary mean-field equation of state. For a ferromagnet, the latter has the usual Landau 
form. For a ferrimagnet, the equation of state depends on details of the magnetic order. It can be
complicated and describe several different phases, see,
e.g., Ref.\ \onlinecite{Plumer_Caille_Hood_1989}. However, generically the first phase encountered
as one approaches from the paramagnetic state is entered via a second-order transition. After
minimizing $f_0$ and expressing $n$ in terms of $m$ one thus has again an ordinary mean-field
equation of state given by
\be
h = r\,m + u\,m^3 + O(m^5)\ ,
\label{eq:2.24}
\ee
where $h$ is an external magnetic field in the 3-direction, $u>0$, and the transition occurs at 
$r=0$.\cite{Landau_theory_footnote}
In Appendix \ref{app:A} we recall a very simple model that leads to this result.
The second term on the right-hand side
of Eq.\ (\ref{eq:2.23a}) gives an additional contribution to the equation of state, which then reads
\bea
h &=&  r\,m + u\,m^3 - 64\,m\,c^2\gamma_{\text{t},1}\gamma_{\text{t},2}\,
\nonumber\\
&& \hskip 20pt \times \frac{1}{V}{\sum_{\bm k}}^{\prime} T\sum_{n=1}^{\infty} 
   \frac{\Omega_n^2\,\left(\varphi^{(3)}({\bm k},\Omega_n)\right)^2\,
   \varphi^{-4}({\bm k},\Omega_n)}{N({\bm k},\Omega_n;m)}\ .
\nonumber\\
\label{eq:2.25}
\eea
This is the desired generalized mean-field equation of state which takes into account the coupling
of the order parameter to the fermionic soft modes.

\subsection{Discussion of the Generalized Mean-Field Equation of State}
\label{subsec:II.E}

With some effort the integrals in Eqs.\ (\ref{eq:2.23b}) and (\ref{eq:2.25}) can be explicitly performed.
However, the salient points can be seen by simple scaling considerations and dimensional analysis.
Equations (\ref{eq:2.11}) and (\ref{eq:2.13}) imply that the frequency $\Omega_n$ scales as the wavenumber $k$,
$\Omega_n \sim k$, and that $\varphi({\bm k},\Omega_n) \sim 1/k \sim 1/\Omega_n$, which also can be
seen explicitly from Eqs.\ (\ref{eqs:2.14}). Equation (\ref{eq:2.15}) implies that 
$\varphi^{(3)}({\bm k},\Omega_n) \sim 1/k^2 \sim 1/\Omega_n^2$. Equation (\ref{eq:2.23c}) then shows
that there is a length scale $L_m$, or a corresponding frequency scale $\omega_m$, that scales as
$L_m \sim 1/\omega_m \sim 1/m$. If one attempts to expand $\Delta f(m)$, Eq.\ (\ref{eq:2.23b}),
in powers of $m$ at $T=0$, then nonanalyticities will occur at next-to-leading order for all $d\leq 3$.

An alternative way to describe this mechanism is to say that of the three soft fermionic 
spin-triplet excitations, Eq.\ (\ref{eq:2.17a}) with $r=s=0,3$ and $i=j=1,2,3$, two (namely, the 
ones transverse to the order parameter
direction) acquire a mass due to the coupling between the
fermions and the order parameter $m$, as can be seen explicitly from Eq.\ (\ref{eq:2.22b}). 
This acquisition of a mass by a generic soft mode due the spontaneous breaking of a continuous
symmetry is an example
of the Anderson-Higgs mechanism,\cite{Anderson_1963, Higgs_1964a, Higgs_1964b} 
even though the broken symmetry in this case is not a gauge symmetry,
see the discussion in Sec.\ \ref{subsec:III.A}. It implies in turn that
the free energy is a nonanalytic function of $m$.

At nonzero temperatures the singularities are cut off by $T$ according to $m \sim T$. That is, a
crossover occurs from $m$-scaling to $T$-scaling when the Zeeman splitting is comparable to the
temperature, or the thermal length scale $L_T \propto 1/T$ is comparable to the magnetic length
scale $L_m$ mentioned above. Taking into account the sign of $N$, Eq.\ (\ref{eq:2.23c}), one finds
schematically, for $1<d<3$,
\bse
\label{eqs:2.26}
\be
\Delta f(m) = -v\,m^2(m^2 + T^2)^{(d-1)/2}\ ,
\label{eq:2.26a}
\ee
and for $d=3$
\be
\Delta f(m) = \frac{v}{8}\,m^4\ln(m^2+T^2)\ ,
\label{eq:2.26b}
\ee
\ese
where $v>0$ is a positive constant.

\begin{figure}[t]
\vskip -0mm
\includegraphics[width=8.0cm]{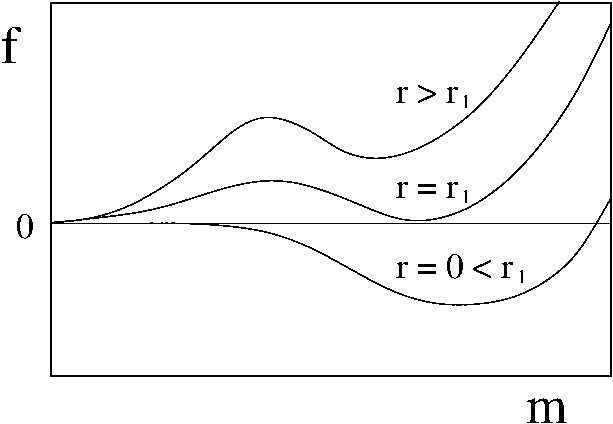}
\caption{Schematic sketch of the free energy for three values of the parameter $r$. The first-order
 transition occurs at $r = r_1 > 0$. It pre-empts the second-order transition of Landau theory
 which would occur at $r=0$.}
\label{fig:2}
\end{figure}

The most important aspects of this result, as far as the order of the transition is concerned, are the
sign of $v$ and the power of $m$ at $T=0$. For all $d\leq 3$ there is a negative term in the free
energy that dominates the $m^4$ in the Landau free energy and hence necessarily leads to a
first-order transition. Another way to see this is by expanding $\Delta f(m)$, Eq.\ (\ref{eq:2.26a}),
in powers of $m$ for $T>0$. The leading term is proportional to $-m^4/T^{3-d}$. That is, there
is a {\em negative} $m^4$ term whose prefactor diverges as $T\to 0$ for all $d\leq 3$, which implies
that there will be a tricritical point at some temperature. The free energy for three different values
of $r$ is plotted schematically in Fig.\ \ref{fig:2}.
For this schematic free energy, the equation of state in the case $d=3$, for which many
experimental results exist, takes the form
\bea
h &=& r\,m + \frac{v}{2}\,m^3 \ln(m^2+T^2) \hskip 80pt
\nonumber\\
&& + m^3\left(u + \frac{v}{4}\,\frac{m^2}{m^2+T^2}\right)\ .\quad (d=3)
\label{eq:2.27}
\eea
Also of interest is the other physical dimensionality, $d=2$, where the equation of state reads
\bea
h &=& r\,m - 2v\,m(m^2+T^2)^{1/2} \hskip 80pt
\nonumber\\
&&+ m^3\left(u - \frac{v}{(m^2+T^2)^{1/2}}\right)\ .\quad (d=2)
\label{eq:2.28}
\eea
Here the analyticity is stronger than in the $3$-$d$ case, with a negative $m^2$-term in the
equation of state at $T=0$. This is particularly interesting in the case of Ising magnets, which
display long-range order in $d=2$ even at $T>0$. The case of Heisenberg and XY magnets,
which do not show true long-range order in $d=2$ except at $T=0$, is more complicated.

These are the same results that were obtained using a more phenomenological theory of the
fermionic soft modes in Ref.\ \onlinecite{Kirkpatrick_Belitz_2003b}. They were discussed
extensively in that reference, as well as in Refs.\ \onlinecite{Belitz_Kirkpatrick_Vojta_1999}
and \onlinecite{Belitz_Kirkpatrick_Rollbuehler_2005}. There is no need to repeat this discussion
here, and the salient features are summarized by the schematic phase diagram shown in 
Fig.\ \ref{fig:1}. The important conclusion of the current paper is that the validity of these
results, in addition to itinerant Heisenberg ferromagnets, extends to
metallic ferromagnets where the magnetism is not due to the conduction electrons, to metallic
ferromagnets in the XY or Ising universality class, and also to metallic ferrimagnets. The only condition 
is that the conduction electrons are not subject to strong spin-symmetry breaking effects such as
magnetic impurities. We note in passing that an interesting system is provided by the easy-plane 
ferromagnet URhGe, where an in-plane magnetic field transverse to the magnetization has been
used to tune the transition, access the tricritical point, and map out the tricritical wings.\cite{Yelland_et_al_2011} This
situation requires a refinement of the theory presented above, which will be reported
elsewhere.\cite{us_tbp}

\section{Discussion, and Conclusion}
\label{sec:III}

We now discuss our results, before concluding with a summary.

\subsection{The mechanism behind the first-order transition}
\label{subsec:III.A}

The mechanism that leads to the first-order transition discussed in Sec.\ 
\ref{subsec:II.E} is precisely analogous to the fluctuation-induced first-order
transition discussed in Ref.\ \onlinecite{Halperin_Lubensky_Ma_1974} for
the BCS-superconductor transition and the nematic-to-smectic-A transition
in liquid crystals. An important physical ingredient is
an underlying ``generic" soft mode, i.e., one that is not related to
the phase transition in question, but couples to the order parameter. In
the case of liquid crystals this soft mode is the nematic Goldstone mode,
in the case of superconductors, the vector potential, in the present case,
the spin-triplet particle-hole excitation. At the transition of interest,
this soft mode acquires a mass that is given in terms of the nonzero
expectation value of the order parameter. This general mass-generating
mechanism was first pointed out by Anderson, and is now known as the
Anderson-Higgs mechanism.\cite{Anderson_1963, Higgs_1964a, Higgs_1964b}
This coupling of the order parameter to underlying soft modes leads to 
a non-analytic term in the Landau free energy that is dominant over the
usual quartic term and has a negative sign, leading to a first-order
transition. It should be stressed that this is only one way to realize a
fluctuation-induced first-order transition; another one, for instance, is
realized by a $\phi^4$-theory with a cubic anisotropy.\cite{Wallace_1973}
The current realization is analogous to the case of scalar electrodynamics
studied by Coleman and Weinberg in a particle-physics 
context.\cite{Coleman_Weinberg_1973} It is also worthwhile noting that
the analogy between superconductors on one hand, and liquid crystals
and quantum magnets on the other, breaks down in the ordered phase.
In the former case, the Goldstone mode gets absorbed into the
longitudinal component of the vector potential, which is massive, and there
is no soft mode in the ordered phase. In the latter, there are Goldstone modes
in the ordered phases, namely, a ``smecton'' with an anisotropic dispersion
relation in the smectic-A phase (Ref.\ \onlinecite{DeGennes_Prost_1993}, see
also Ref.\ \onlinecite{Kirkpatrick_Belitz_2009a}) and magnons
in the magnetic phase.

\subsection{Universality of the first-order transition, and the validity of the
generalized mean-field theory}
\label{subsec:III.B}

Experimentally, all examples of clean low-T$_{\text{c}}$ ferromagnets
(for disordered systems, see below; ferrimagnets so far have not been systematically
studied from this point of view) show a first-order transition if the Curie
temperature is suppressed far enough. There is not a single example of a
quantum critical point in zero magnetic field. While this is consistent with
the generalized mean-field theory theory presented in Sec.\ \ref{sec:II}, it
is somewhat surprising when compared with the case of liquid crystals,
where an analogous theory also predicts a first-order transition. In this
case, in stark contrast to that of quantum magnets, the observed transition
is usually of second order, and only recently have examples of a (weakly)
first-order transition been found.\cite{Yethiraj_Mukhopadhyay_Bechhoefer_2002}
These observations beg the question whether in the case of quantum magnets
the generalized mean-field approximation is more generally valid than in
classical systems.

To discuss this point, we first observe that we have made three approximations
to treat the action given by Eq.\ (\ref{eq:2.1a}). 
First, we have integrated out the fermionic massive modes in a saddle-point approximation
that respects the Ward identity that governs the soft-mode structure of the 
system.\cite{Kirkpatrick_Belitz_2012, Belitz_Kirkpatrick_2012} Second, we have kept 
the soft fermionic degrees of
freedom only to Gaussian order in the soft modes $q$. Third, we have treated the order
parameter in a mean-field approximation. These approximations are not independent of
one another, and the first two simplifications do not constitute any additional
approximation over and above the last one. This can be seen as follows.

The mean-field approximation for the order parameter means that the fermionic degrees
of freedom describe an interacting electron system that is spin-polarized by the coupling
to the homogeneous magnetization, which acts as an effective external magnetic field.
The state of the fermionic subsystem is thus described by a stable Fermi-liquid fixed
point. Corrections to the fermionic soft-mode action due to massive degrees of freedom
are irrelevant with respect to this fixed point by at least one-half power of frequency or
wavenumber in all dimensions, and thus cannot change the properties of 
system.\cite{Belitz_Kirkpatrick_1997} Similarly, only the terms quadratic in $q$ contribute
to the fixed-point action; all higher-order terms are irrelevant by power counting. Keeping terms
of higher order in $q$ will therefore renormalize the parameters of the theory, but it
cannot change its structure. In particular, it cannot change the sign of the term in the
equation of state, Eqs.\ (\ref{eq:2.27}, \ref{eq:2.28}), that is due to the soft fermionic fluctuations and
leads to the first-order transition. 

This leaves the mean-field approximation for the order parameter to be discussed.
If the first-order transition at $r=r_1$ occurs far from the second-order transition
at $r=0$ that is pre-empted by it (see Fig.\ \ref{fig:2}), then order-parameter fluctuations 
are negligible and the results of the generalized mean-field theory are qualitatively correct. 
If, however, the first-order transition occurs close to the putative second-order one, i.e.,
if the minimum in the free energy in Fig.\ \ref{fig:2} is very shallow, then it is less clear
whether order-parameter fluctuations can be neglected. One key difference between
classical liquid crystals and quantum magnets is that in the former case, the system is
below the upper critical dimension $d_{\text{c}}^+=4$ for the (unrealized) phase
transition that would occur in the absence of any coupling between the smectic
order parameter and the nematic soft modes. In contrast, the quantum magnetic
systems are above the corresponding upper critical dimension $d_{\text{c}}^+=1$
that follows from Hertz theory, and even with that coupling taken into account, ordinary
mean-field theory becomes exact, as far as the description of the phase transition is
concerned, for $d>3$.\cite{Belitz_Kirkpatrick_Vojta_2005} This strongly suggests
that order-parameter fluctuations are of much less importance in the case of quantum
magnets, and it provides a possible explanation of the fact that the observed transition is
universally of first order.

Irrespective of these observations, the role of order-parameter fluctuations
in quantum magnets is a topic that warrants additional work. For the case
where the magnetism is not produced by the conduction electrons, this
will require an action that properly describes localized magnetic moments
and their fluctuations, e.g., the one given in Ref.\ \onlinecite{Read_Sachdev_1995}. For itinerant
magnets, i.e., if the magnetism is due to the conduction electrons themselves,
the theory developed in Sec.\ \ref{sec:II} will apply, but the order-parameter
fluctuations and the fermionic excitations both need to be kept, along the
lines of the phenomenological theory of Ref.\ \onlinecite{Kirkpatrick_Belitz_2003b}.
The latter reference gave a scenario that can lead to a second-order transition
in the magnetic case. It would also be interesting to experimentally study
quantum ferromagnets or ferrimagnets in $d=2$, where order-parameter fluctuations
will be stronger than in $d=3$. 

\subsection{The effects of quenched disorder}
\label{subsec:III.C}

So far we have discussed the case of clean or pure magnets. Impurities,
modeled by quenched disorder, have important effects that are both needed
to understand experimental observations in certain systems, and to predict
effects that can serve to ascertain that the first-order transition in pure
samples is indeed due to the posited mechanism.

Quenched disorder changes the soft-mode spectrum of the fermions.
It gives the ballistic soft modes that are represented by Eqs.\ (\ref{eqs:2.17})
as mass, and leads to new soft modes that are diffusive. In the context of
the current theory, this change has two principal effects. First, it cuts off
the nonanalyticity in the clean equation of state, Eqs.\ (\ref{eq:2.27}, \ref{eq:2.28}).
Second, it leads to a new nonanalytic term in the equation of state that has
the opposite sign and whose prefactor vanishes in the clean 
limit.\cite{Belitz_Kirkpatrick_Vojta_2005} The resulting schematic generalized Landau
theory has been discussed in Ref.\ \onlinecite{Belitz_Kirkpatrick_Vojta_1999}. A more
detailed model discussion that allows for semi-quantitative predictions of the
effects of disorder will be presented elsewhere;\cite{us_tbp} here we just present the most
pertinent aspects of such a model calculation. A good representation of the mean-field 
equation of state for realistic values of the magnetization, the temperature, and the 
disorder, is
\bea
h &=& r\,m + \frac{v^{1/4}}{4(\kF\ell)^{3/2} }\ \frac{m^3}{m^{3/2} + (bT)^{3/2}}
\nonumber\\
&&+ \frac{v}{2}\,m^3\,\ln\left[c\,m^2 + (1/\kF\ell + bT)^2\right]
   + u\,m^3\ ,\qquad
\label{eq:3.1}
\eea
which generalizes Eq.\ (\ref{eq:2.27}) in the presence of quenched disorder. Here
the magnetic field $h$ and the temperature $T$ are measured in units of the 
Fermi energy $\epsilon_{\text F}$ and the Fermi temperature $T_{\text{F}}$, respectively, 
and the magnetization $m$ is measured in units of the
conduction electron density (we put $\mu_{\text{B}}=1$). The dimensionless coupling constant $v$ is proportional to the
fourth power of the effective spin-triplet interaction amplitude of the
conduction electrons. It is a measure of how strongly correlated the conduction
electrons are, and it is bounded above by a stability criterion that requires
$v \alt 0.5$. $\kF$ is the Fermi wave number of the conduction electrons,
and $\ell$ is the elastic mean-free path. Within a Drude model, and for good
metals, one has approximately $\kF\ell \approx 1,000/(\rho_0/\mu\Omega{\text cm})$,
with $\rho_0$ the residual electrical resistivity. $c$ and $b$ are dimensionless
constants that are equal to $c = 1/45$ and $b = 3\pi$ in a model calculation.\cite{us_tbp}
The second factor in the second term on the right-hand side is a reasonable representation,
for realistic parameter values, of a more complicated scaling function 
\be
m^{3/2}\,g(\kF\ell\,m,bT/m) \ \approx \frac{m^3}{m^{3/2} + (bT)^{3/2}}
\label{eq:3.2}
\ee
that depends on the disorder in addition to the temperature, and we have dropped the last term in 
Eq.\ (\ref{eq:2.27}) from Eq.\ (\ref{eq:3.1}) since one generically expects $v\ll u$. 

At $T=0$, and in a clean system, Eq.\ (\ref{eq:3.1}) yields a first-order transition at
$r_1 = v\,m_1^2/4$, where the magnetization discontinuously jumps from $m=0$
to $m = m_1 = e^{-(1+2u/v)/2}$. With $u \approx 0.14$ and $v \approx 0.02$ this yields
$m_1 \approx 4\times 10^{-3}$, which is reasonable for a weak ferromagnet.
Similarly, there is a tricritical temperature given by $T_{\text{tc}}/T_{\text{F}} = (1/b) \exp(-u/v)$;
with the same parameter values this yields $T_{\text{tc}}/T_{\text{F}} \approx 10^{-4}$, 
or $T_{\text{tc}} \approx 10\,\text{K}$ for $T_{\text{F}} = 100,000\,\text{K}$, which
is also reasonable. This tricritical point gets destroyed by quenched disorder on the
order of $\kF\ell \approx bT_{\text{tc}}/T_{\text{F}} \approx 1,000$, or a residual
resisitivity on the order of $\rho_0 \approx 1 \mu\Omega{\text{cm}}$. At this point
the second term on the right-hand side of Eq.\ (\ref{eq:3.1}) is still very small, and
the critical behavior at the resulting quantum critical point is given by ordinary
mean-field exponents except extrely close to the transition, where it crosses over
to the critical behavior derived in Ref.\ \onlinecite{Belitz_Kirkpatrick_1997}. For instance,
in this asymptotic
region the critical exponents $\beta$ and $\delta$, defined by $m(h=0) \propto \vert r\vert^{\beta}$
and $m(r=0) \propto h^{1/\delta}$, respectively, are given by $\beta = 1/2$ and $\delta = 3/2$,
as opposed to the mean-field values $\beta = 1/2$ and $\delta = 3$.
Only for substantially larger values of the disorder, $\rho_0 \approx 100 \mu\Omega{\text{cm}}$
with the above parameters, does the asymptotic critical behavior extend over a
sizeable range of $r$ values (up to $\vert r\vert \approx 0.01$). This observation 
explains why an experiment 
on Ni$_x$Pd$_{1-x}$, which shows a ferromagnetic transition at a very small value
of $x$ ($x \approx 0.025$) corresponding to weak disorder, found mean-field
exponents consistent with Hertz theory,\cite{Nicklas_et_al_1999} whereas 
Bauer et al.\cite{Bauer_et_al_2005} found non-mean-field exponents, at least
some of which were consistent with Ref.\ \onlinecite{Belitz_Kirkpatrick_1997},
in URu$_{2-x}$Re$_x$Si$_2$, where the ferromagnetic transition occurs at
$x \approx 0.15$ with the residual resistivity on the order of 
$\rho_0 \approx 100 \mu\Omega{\text{cm}}$.\cite{Butch_Maple_2010}

\subsection{Conclusion}
\label{subsec:III.D}

In conclusion, we have extended a previous theory of quantum ferromagnets in
several important ways. We have shown that the mechanism that leads to the
paramagnet-to-ferromagnet transition at low temperature in $d=3$ and $d=2$ to be generically of
first order, which was first reported in Ref.\ \onlinecite{Belitz_Kirkpatrick_Vojta_1999},
is valid in anisotropic ferromagnets, in ferrimagnets, and in metallic ferromagnets
where the conduction electrons are not the source of the magnetization, in 
addition to the case of isotropic itinerant ferromagnets originally considered.
This explains why the low-temperature transition is observed to be of first
order in highly anisotropic ferromagnets, and it much expands the class of
materials for which this phenomenon is predicted.
For clean magnets, an effective theory of soft fermionic modes recently
developed in Ref.\ \onlinecite{Belitz_Kirkpatrick_2012} has provided a technical
basis that improves on the phenomenological theory of Ref.\ \onlinecite{Kirkpatrick_Belitz_2003b}.
In the presence of quenched disorder, the theory allows for a semi-quantitative
description of the suppression and ultimate destruction of the tricritical point. A sizeable
range of disorder exists where the observable critical behavior is predicted to be
mean-field like, whereas for very large disorder the asymptotic critical region,
which is characterized by non-mean-field Gaussian critical exponents, expands
and eventually eliminates the mean-field region.

\appendix
\section{A simple mean-field model of a ferrimagnet}
\label{app:A}
Here we recall a very simple mean-field model of the transition from a paramagnet to
long-range ferrimagnetic order.\cite{Kittel_1996} Consider a one-dimensional chain
of alternating magnetic moments $\mu_a$, $\mu_b$ that are antiferromagnetically
coupled. Weiss theory assumes that the $a$-moments and $b$-moments are subject to effective
magnetic fields 
\bse
\label{eqs:A.1}
\bea
B_a &=& -\lambda\,M_b
\label{eq:A.1a}\\
B_b &=& -\lambda\,M_a\ ,
\label{eq:A.1b}
\eea
\ese
respectively, where $\lambda > 0$. The magnetizations $M_{a,b}$ are given by the Brillouin expressions
\bse
\label{eqs:A.2}
\bea
M_a &=& \nu\,\mu_a\,\tanh(\mu_a\,H/T + \mu_a\,B_a/T)\ ,
\label{eq:A.2a}\\
M_b &=& \nu\,\mu_b\,\tanh(\mu_b\,H/T + \mu_b\,B_b/T)\ .
\label{eq:A.2b}
\eea
\ese
Here $H$ is an external magnetic field, $T$ is the temperature, and $\nu$ is the number of magnetic moments
of each species. If one defines reduced magnetic fields $h_{a,b} = H/\nu\mu_{a,b}\lambda$,
a reduced temperature $t = T/\nu\mu_a\mu_b\lambda$, and reduced moments
$m_{a,b} = M_{a,b}/\nu\mu_{a,b}$, then one sees that the Weiss mean-field equations
(\ref{eqs:A.1}, \ref{eqs:A.2}) have a solution $m_a = -m_b = \tilde{m}$, where $\tilde{m}$ is
the solution of the usual mean-field equation of state
\be
h = r{\tilde m} + {\tilde m}^3/3 + O({\tilde m}^5)\ ,
\label{eq:A.3}
\ee
where $r = t-1$. This simple model thus describes a transition at $t=1$ to ferrimagnetic
order where the homogeneous magnetization is given by $m = M_a + M_b = \nu(\mu_a-\mu_b){\tilde m}$
and the staggered magnetization $n = M_a - M_b = \nu(\mu_a+\mu_b){\tilde m}$ is
proportional to $m$.

\acknowledgments
We gratefully acknowledge discussions and correspondence with Greg Stewart, Jeff Lynn, and Nick Butch. 
This work was supported by the National Science Foundation under Grant Nos. DMR-09-29966, and 
DMR-09-01907.


\begin{thebibliography}{37}
\bibitem[{dim()}]{dimensionality_footnote}
\bibinfo{note}{In these and related statements we denote by $d$ the spatial
  dimensionality of the system; spin space is always considered to be
  three-dimensional.}
\bibitem[{MFT()}]{MFT_footnote}
\bibinfo{note}{This is true within the framework of the generalized mean-field
  theory whose validity is discussed in Sec.\ \ref{subsec:III.B}. If
  order-parameter fluctuations are important it is possible that the behavior
  is less universal.}
\bibitem[{Lan()}]{Landau_theory_footnote}
\bibinfo{note}{In any given material it is of course possible that the
  parameter $u$ in the Landau theory is negative, leading to a first-order
  transition even within Landau theory. However, this will not be the case
  generically, whereas the generalized mean-field theory predicts a generic
  first-order transition.}
\end{thebibliography}

\end{document}